 \documentclass[12pt]{article} 
 \usepackage{color} 
 \newcommand{\comma}{\;\; ,}
 \newcommand{\period}{\;\; .}
 \newcommand{\eq}{\; = \;}
 \newcommand{\sep}{\;\; , \;\;}
 \newcommand{\be}{\begin{equation}}
 \newcommand{\bd}{\begin{displaymath}}
 \newcommand{\ee}{\end{equation}}
 \newcommand{\ed}{\end{displaymath}}
 \newcommand{\ba}{\begin{eqnarray}}
 \newcommand{\ea}{\end{eqnarray}}
 
 \newcommand{\Baxter}{B~}
 \newcommand{\zp}{p}
 \newcommand{\zq}{q}
 \newcommand{\zpb}{\overline{p}}
 \newcommand{\zqb}{\overline{q}}
 \newcommand{\zpt}{\hat{p}}
 \newcommand{\zqt}{\hat{q}}
 \newcommand{\minus}{\! - \!}
 
 \newcommand{\plus}{\! + \!}
 \newcommand{\del}{\!\!\!\!\!}
 \renewcommand{\i}{{\rm i}}
 \newcommand{\e}{{\rm e}}

 \title{Free energy of the three-state $\tau_2(t_q)$ model as
 a product of elliptic functions}

 \author{ R.J. Baxter, F.R.S. \\
 {\protect \small  Mathematical
 Sciences Institute}\\
 {\protect  \small The Australian National University,
 Canberra, A.C.T. 0200,
  Australia, (\small e-mail: none) }}

 \date{\protect \small Tuesday 14  February 2006}

\begin{document}


 \maketitle

 \abstract{We show that the free energy of the three-state 
 $\tau_2(t_q)$ model can be expressed as products of Jacobi elliptic
 functions, the arguments being those of an hyperelliptic
 parametrization of the associated chiral Potts model. This is the 
 first application of such a parametrization to the $N$-state chiral 
 Potts free energy problem for $N > 2$.}

{\bf Keywords: Statistical mechanics, lattice models, free energy, chiral Potts model,
$\tau_2$ model}






 \begin{center}
 \section*{1. Introduction}
 \end{center}
 \setcounter{equation}{0}
 \renewcommand{\theequation}{1.\arabic{equation}}


 In the field of solvable models in statistical mechanics, the 
 $N$-state two-dimensional chiral Potts model has proved particularly 
 challenging. This is despite the fact that for $N=2$ it reduces to 
 the Ising model, the free energy of which was calculated by Onsager 
  (1944).

 The model was developed by Howes, Kadanoff and den Nijs 
 (1983), von Gehlen and Rittenberg (1985), Au-Yang {\it et al} (1987)
 and McCoy, Perk and Tang (1987). It was first fully defined as a 
 general $N$-state lattice model by Baxter, Perk \& Au-Yang (1988), 
 who showed that it satisfied the star-triangle relations. These 
 relations define ``rapidity'' variables $p, q$ such that the  
 Boltzmann weights are functions $W_{pq}$ of $p$ and $q$.

 Many of the citations herein are to earlier papers by the author: in 
 these we shall abbreviate ``Baxter'' to ``B''.

 With previously solved models, such as the hard-hexagon 
 model (\Baxter 1980), the calculation of the free energy was 
 straightforward once one had obtained the star-triangle relation and 
 rapidities for that model. Typically, it turned out that there was
 a transformation in terms of Jacobi elliptic functions such that
 $W_{pq}$ depended on $p, q$ only via their difference $u = q-p$.
 The free energy is $- k_B T \log \kappa_{pq}$, where $\kappa_{pq}$
 is the partition function per site. This $\kappa_{pq}$ must also
 be a function  $\kappa (u)$ of $u$, and there were always
 ``rotation'' and ``inversion'' relations (\Baxter 1982a, 1982b), of 
 the form
 \bd \kappa (u) \eq  \kappa (\lambda - u) \ed
 \be \label{inv}
 \kappa (u) \, \kappa (-u) \eq r(u) \comma \ee 
 where $\lambda$ is a positive real constant (the 
 ``crossing parameter'') and $r(u)$ a known meromorphic function.
 For $u$ real and $0 < u < \lambda$ the Boltzmann weights are 
 real and positive, and 
 one can develop low-temperature series expansions in the usual way
 that indicate that $\kappa(u)$ is analytic, non-zero and bounded
 in the vertical strip $0 \leq \Re (u) \leq \lambda$. The equations
 (\ref{inv}) then determine $\kappa(u)$ and one can solve them by 
 Fourier transforms or other methods.

 The $N > 2$ chiral Potts model, however, does {\em not} have the
 ``rapidity difference property'', i.e. there is no 
 transformation that takes $W_{pq}$ to a function only of $q-p$.
 This makes the calculation of $\kappa_{pq}$ much more difficult, 
 and it was not till 1990 that an explicit  result was obtained 
 (\Baxter 1990, 1991a). The calculation of the 
 spontaneous magnetizations (order parameters) is even harder, and 
 was not accomplished until 2005 (\Baxter 2005a, 2005b).

 The  calculation of $\kappa_{pq}$ proceeds in two stages. First one 
 calculates the partition function per site $\tau_2(p,q)$ of an 
 associated ``$\tau_2(t_q)$ model'', which is closely related to the
 superintegrable case of the chiral Potts model. Then one uses this
 result to calculate  $\kappa_{pq}$. 

 In the calculation of the order parameters, we used the 
 fact that certain functions $G_{p,Vp}(r)$, ${\cal S}(p)$ are similar
 to $\tau_2(p,q)$. In fact they are ratios of special cases of 
 $\tau_2(p,q)$. 

 Although the main task, namely the calculation of the free energy
 and order parameters of the chiral Potts model, has been 
 accomplished, it remains disappointing that one has no elegant 
 parametrization in terms of elliptic functions, as one has for models 
 with the difference property. These parametrizations explicitly 
 exhibit the poles and zeros of $\kappa_{pq}$ on an extended
 $p,q$ Riemann surface. There is a parametrization of the 
 rapidities and Boltzmann weights of the $N$-state chiral Potts model 
 in terms of hyperelliptic functions (\Baxter 1991b), but these have 
 $N-1$ arguments that are related to one another in a complicated way, 
 and until recently they have not been found to be particularly 
 useful.

 However, we have now found that for $N=3$ the function ${\cal S}(p)$ 
 mentioned above can be expressed as a ratio of generalized elliptic 
 functions of these arguments (\Baxter 2006). The obvious question is 
 whether $\tau_2(p,q)$ can be similarly expressed. We show here that 
 the answer is yes. The functions that occur are the same as that 
 appear in other solvable models, notably the Ising model.





\vspace{5mm}
\begin{center}
{\section*{2. The function $\tau_2(p,q)$}}
\end{center}
\vspace{-6mm}
\setcounter{equation}{0}
 \renewcommand{\theequation}{2.\arabic{equation}}


 For the chiral Potts model, the rapidity $p$ can be thought of as the 
 set of variables $p = \{ x_p,y_p,\mu_p, t_p \}$, related to one 
 another by
 \ba \label{prlns}   t_p = x_p y_p \sep & & \del \del 
 x_p^N + y_p^N = k(1+x_p^N y_p^N) 
 \comma \nonumber \\
 && \\
 k x_p^N = 1-k'/\mu_p^N 
 & , & k y_p^N = 1-k'\mu_p^N \period \nonumber \ea
 Here $k, k'$ are real positive constants, satisfying
 \be \label{kkp}
 k^2 + {k'}^2 \eq  1 \period \ee
 In terms of the $a_p, b_p, c_p, d_p$ of (Baxter, Perk 
 \& Au-Yang 1988; \Baxter 1991b),
 $x_p = a_p/d_p,\,  y_p = b_p/c_p, \, \mu_p = d_p/c_p$.

 There are various automorphisms or maps that take one set 
 $\{ x_p,y_p,\mu_p, t_p \}$ to another set satisfying the same 
 relations (\ref{prlns}). Four that we shall use are:
 \ba \label{autos}
 R:  \{ x_{Rp},y_{Rp},\mu_{Rp},t_{Rp} \} & = &
 \{ y_p,\omega x_p,1/\mu_p, \omega t_p \} \comma \nonumber \\
 U:  \{ x_{Up},y_{Up},\mu_{Up},t_{Up} \} & = & 
 \{ x_p^{-1},y_p^{-1},\omega^{-1/2} x_p \mu_p/y_p,  
 t_p^{-1} \} \comma \nonumber \\
 V:  \{ x_{Vp},y_{Vp},\mu_{Vp},t_{Vp} \} & = & 
 \{ x_p,\omega y_p, \mu_p, \omega t_p \} \comma  \\
 M:  \{ x_{Mp},y_{Mp},\mu_{Mp},t_{Mp} \} & = & 
 \{ x_p, y_p, \omega \mu_p,  t_p \} \comma \nonumber \ea
 where \bd
 \omega \eq \e^{2 \pi \i/N} \period \ed
 They satisfy
 \bd 
 R V^{-1} R = V  \sep M R M = R   \comma \ed
 \be \label{Veqns} U^2 M =  V^N = M^N =  1   \period \ee
 Here   $U = RSV$, $S$ being the operator $S$ used in (Baxter, Perk 
 \& Au-Yang 1988, \Baxter 2006).

We take $\mu_p$ to be outside the unit circle, so
 \be \label{mupr}
 |\mu_p | > 1 \period \ee 
 Then we can specify
 $x_p$ uniquely by requiring that
 \be \label{xpr}
 -\pi/(2 N) < \arg (x_p ) < \pi/(2 N) \period \ee


 \setlength{\unitlength}{1pt}
 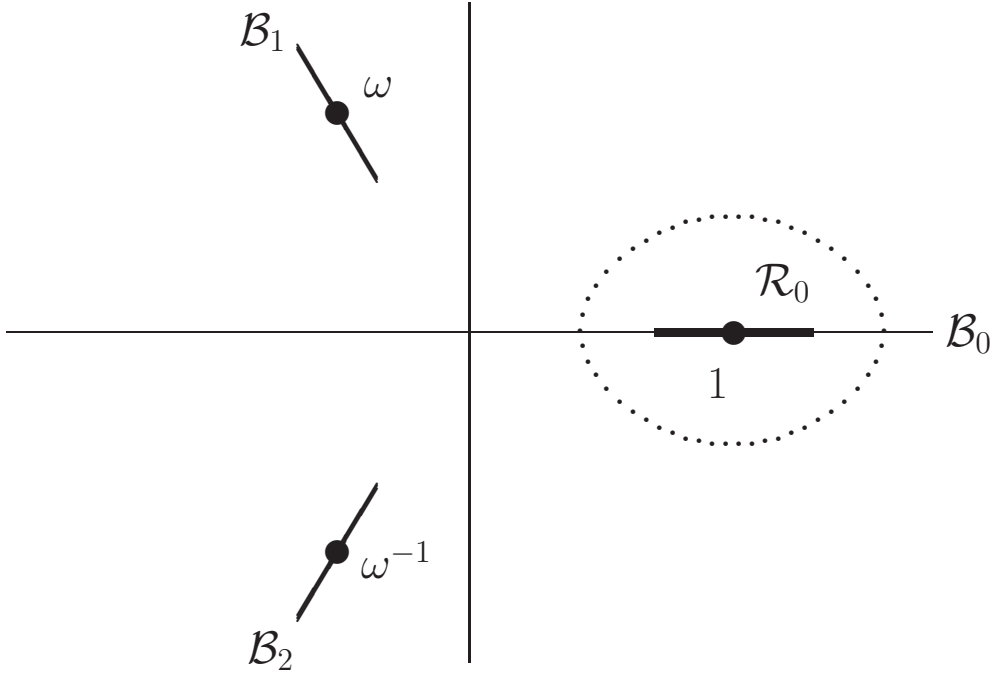
\begin{figure}[hbt]
 \begin{picture}(420,260) (0,0)

 \put (50,125) {\line(1,0) {350}}
 \put (225,0) {\line(0,1) {250}}

 \put (325,125)  {\circle*{9}}
 \put (175,208)  {\circle*{9}}
 \put (175,42)  {\circle*{9}}

 \put (405,120)  {\Large  {${\cal B}_0$}}
 \put (138,234)  {\Large  {${\cal B}_1$}}
 \put (141,0)  {\Large  {${\cal B}_2$}}

 \put (315,100)  {\Large 1}
 \put (185,214)  {\Large $\omega$}
 \put (184,32)  {\Large $\omega^{-1}$}

 \put (333,139) {{\Large {${\cal R}_0$}}}

 \thicklines
 \put (295,124) {\line(1,0) {60}}
 \put (295,125) {\line(1,0) {60}}
 \put (295,126) {\line(1,0) {60}}

 \put (265,125) {\bf .}
 \put (266,131) {\bf .}
 \put (268,137) {\bf .}
 \put (271,142) {\bf .}
 \put (275,147) {\bf .}
 \put (280,152) {\bf .}
 \put (285,156) {\bf .}
 \put (290,159) {\bf .}
 \put (295,162) {\bf .}
 \put (300,164) {\bf .}
 \put (305,166) {\bf .}
 \put (310,167) {\bf .}
 \put (315,168) {\bf .}
 \put (320,168) {\bf .}
 \put (325,168) {\bf .}
 \put (330,168) {\bf .}
 \put (335,167) {\bf .}
 \put (340,166) {\bf .}
 \put (345,164) {\bf .}
 \put (350,162) {\bf .}
 \put (355,159) {\bf .}
 \put (360,156) {\bf .}
 \put (365,152) {\bf .}
 \put (370,147) {\bf .}
 \put (374,142) {\bf .}
 \put (377,137) {\bf .}
 \put (379,131) {\bf .}
 \put (380,125) {\bf .}

 \put (266,119) {\bf .}
 \put (268,113) {\bf .}
 \put (271,108) {\bf .}
 \put (275,103) {\bf .}
 \put (280,98) {\bf .}
 \put (285,94) {\bf .}
 \put (290,91) {\bf .}
 \put (295,88) {\bf .}
 \put (300,86) {\bf .}
 \put (305,84) {\bf .}
 \put (310,83) {\bf .}
 \put (315,82) {\bf .}
 \put (320,82) {\bf .}
 \put (325,82) {\bf .}
 \put (330,82) {\bf .}
 \put (335,83) {\bf .}
 \put (340,84) {\bf .}
 \put (345,86) {\bf .}
 \put (350,88) {\bf .}
 \put (355,91) {\bf .}
 \put (360,94) {\bf .}
 \put (365,98) {\bf .}
 \put (370,103) {\bf .}
 \put (374,108) {\bf .}
 \put (377,113) {\bf .}
 \put (379,119) {\bf .}

 \put (160,16) {\line(3,5) {30}}
 \put (160,17) {\line(3,5) {30}}
 \put (160,18) {\line(3,5) {30}}

 \put (160,234) {\line(3,-5) {30}}
 \put (160,233) {\line(3,-5) {30}}
 \put (160,232) {\line(3,-5) {30}}

 \thinlines

 \ \end{picture}
 \vspace{1.5cm}
 \caption{ The cut $t_p$-plane for $N=3$, showing the three
 branch cuts ${\cal B}_0, {\cal B}_1, {\cal B}_2$ and the 
  approximately circular region ${\cal R}_0$ in which $x_p$ lies 
 when $p \in {\cal D}$.}
 \label{brcuts}
 \end{figure}

 We regard $x_p, y_p, \mu_p^N$ as functions of $t_p$.
 Then $t_p$ lies in a complex plane containing $N$ branch cuts
 ${\cal B}_0, {\cal B}_1, \ldots ,$ $ {\cal B}_{N-1}$ on the lines 
 $\arg (t_p) = 0, 2 \pi /N , \ldots ,$ $ 2 \pi (N-1)/N$,
 as indicated in Fig. \ref{brcuts}, while $x_p$ lies in a 
 near-circular
 region round the point $x_p = 1$, as indicated schematically by the
 region ${\cal R}_0$ inside the dotted curve of Fig. \ref{brcuts}. The 
 variable $y_p$ can lie anywhere in the complex plane {\em except}
 in ${\cal R}_0$ and in $N-1$ corresponding near-circular regions 
 ${\cal R}_1, \ldots ,{\cal R}_{N-1}$ round the  other branch cuts.
 With these choices, we say that $p$ lies in the ``domain'' $\cal D$.

 It can be helpful to consider the low-temperature limit, when 
 $k' \rightarrow 0$. The branch cuts  ${\cal B}_0, 
 {\cal B}_1, \ldots ,$ $ {\cal B}_{N-1}$ and the regions
 ${\cal R}_0,{\cal R}_1, \ldots ,{\cal R}_{N-1}$ then shrink to the 
 points $1, \omega, \ldots, \omega^{N-1}$. If $t_p$ is held fixed, 
 not at $1, \omega, \ldots, \omega^{N-1}$, then
 $x_p \rightarrow 1, y_p \rightarrow t_p$ and $\mu_p^N = O(1/k')$.

 Define $q$ similarly to $p$, also lying in $\cal D$. 
 Then from eq. 39 of (\Baxter 1991a) and eq. 54 of (\Baxter 2003a), 
 the partition function
 per site $\tau_2(p,q) = \tau_2(\mu_p,t_q)$ of the 
 $\tau_2(t_q )$ model is given by
 \be \label{deftau2}
 \log \tau_2(p,q ) \eq \frac{1}{2 \pi} 
 \int_{0}^{2 \pi} \left( \frac{1+  \e^{\i \theta}/\mu_p^N }{1-  
 \e^{\i \theta}/\mu_p^N } \right) 
 \;  \log\left[  \Delta (\theta )  - \omega t_q \right] \, 
 {\rm d} \theta \comma \ee
 where 
 \be \label{defDelta}
 \Delta (\theta ) \eq [(1 - 2 k' \cos \theta +{k'}^2 )/k^2]^{1/N} 
 \period \ee
 We have omitted a factor $y_p^2$ from $\tau_2(p,q )$, and 
 to ensure that this formula is correct for $p, q \in {\cal D}$, we 
 invert $\mu_p, \mu_q$  in (\Baxter 1991a, 2003a).

 This function $\tau_2(p,q)$ satisfies the relations
 \be \label{unvp}
    \tau_2(Vp, q) \eq \tau_2(p, q) \comma \ee
 and
  \be \label{prodtau}
  \prod_{j=0}^{N-1} \tau_2(p,V^j q)  \eq 
 \prod_{j=0}^{N-1} \tau_2(\mu_p,\omega^j t_q) \eq \alpha(p,q) 
 \comma \ee
 where
 \be \label{defalpha}
 \alpha (p,q) \eq \frac{k' 
  (\mu_q^N - \mu_p^{-N})^2} {k^2 \mu_q^N } \eq \frac{
  (x_p^N - y_q^{N})(x_q^N - y_p^N)} {k' \mu_p^N } \period \ee






 \begin{center}
{ \section*{3. The Riemann sheets (``domains'') formed by analytic 
  continuation }}
 \end{center}

\setcounter{equation}{0}
\renewcommand{\theequation}{3.\arabic{equation}}

 We shall want to consider the analytic continuation of certain 
 functions of $t_p$ onto other Riemann sheets, i.e. beyond the domain 
 $\cal D$. We restrict 
 attention to functions that are meromorphic and single-valued in 
 the cut plane of Figure \ref{brcuts}, and similarly for their 
 analytic continuations. Obvious examples  are $x_p, y_p$ and 
 $\tau_2(t_p)$. They are therefore  meromorphic and single-valued 
 on their Riemann surfaces, but we need to know what these 
 surfaces are.

 We start by considering the most general such surface. As a 
 first step,
 allow $\mu_p$ to move from outside the unit circle to inside. Then 
 $t_p$ will cross one of the $N$  branch cuts 
 ${\cal B}_i$ in Figure  \ref{brcuts}, 
 moving onto another Riemann sheet, going back to its original value
 but now with $y_p$ in ${\cal R}_i$.
 Since $y_p$ is thereby confined to  the region near and surrounding
 $\omega^i$, we say that $y_p \simeq \omega^i$. Conversely,
 by $y_p \simeq \omega^i$ we mean that $y_p \in {\cal R}_i$

 We say that $p$ has moved into the  {\em domain}  ${\cal D}_i$
 {\em adjacent} to $\cal D$. There are $N$ such domains
 ${\cal D}_0, {\cal D}_1,$ $\ldots , {\cal D}_{N-1}$.

 Now allow $\mu_p$ to become larger than one, so
 $t_p$ again crosses one of the $N$ branch cuts. Again we require  
 that $t_p$ returns to its original value. If it crosses
 ${\cal B}_i$, then it moves back to the original domain $\cal D$.
 However, if it crosses  another cut ${\cal B}_j$ then 
 $x_p$ moves into ${\cal R}_{j-i}$, 
 and we say that $p$ is now in domain ${\cal D}_{i,j-i}$.

 Proceeding in this way, we build up a Cayley tree of domains.
 For instance, the domain ${\cal D}_{ijk}$ is a third neighbour
 of $\cal D$, linked via the first neighbour ${\cal D}_{i}$ and the
 second-neighbour ${\cal D}_{ij}$, as indicated in Figure
 \ref{seq}. Here $x_p \simeq 1 $ in ${\cal D}$,
 $y_p \simeq \omega^i $ in ${\cal D}_{i}$,
 $x_p \simeq \omega^j $ in ${\cal D}_{ij}$  and 
 $y_p \simeq \omega^k $ in ${\cal D}_{ijk}$. 
 We reject moves that take $p$ back to the domain
 immediately before the last, so $j \neq 0$ and $k \neq i$.
 We refer to the sequence $\{ i,j,k ,\ldots \}$ that define
 any domain as a {\it route}. We can think of it as a sequence of
 points, all with the same value of $t_p$, on the successive
 Reimann sheets or domains.

 The domains $\cal D$, ${\cal D}_{ij}$, ${\cal D}_{ijk\ell}$,...
 with an even number of indices, have $x \simeq \omega^{\ell}$, where
 $\ell$ is the last index. We refer to them as being of even 
 {\it parity} and of {\em type} $\ell$. The domains ${\cal D}_i$, 
 ${\cal D}_{ijk}$,... have $y \simeq \omega^{\ell}$ and are of 
 odd parity and type $\ell$.



 \setlength{\unitlength}{1pt}
 \begin{figure}[hbt]
 \begin{picture}(420,60) (0,30)
 \put (100,25) {\line(1,0) {35}}
 \put (80,20) {\Large ${\cal D}$ }
 \put (150,20) {\Large ${\cal D}_i$}
 \put (175,25) {\line(1,0) {35}}
 \put (220,20) {\Large ${\cal D}_{ij}$}
 \put (250,25) {\line(1,0) {35}}
 \put (295,20) {\Large ${\cal D}_{ijk}$}
 \end{picture}
  \vspace{1.5cm}
 \caption{ A sequence of adjacent domains ${\cal D}, {\cal D}_i, 
 {\cal D}_{ij}, {\cal D}_{ijk}$.}
 \label{seq}
 \end{figure}

 The automorphism that takes a point $p$ in $\cal D$ to 
 a point in  ${\cal D}_i$, respectively, is
 the mapping
 \be
 A_{i} \eq  V^{i-1} \, R V^{-i} \period \ee
 If $p' = A_i \, p$, then
 \be \label{amap}
 x_{p'} = \omega^{-i} y_p \sep y_{p'} = \omega^i x_p \sep t_{p'} = t_p 
 \period \ee
 Because of (\ref{Veqns}), $A_{i+N} = A_i$, so there are $N$ such 
 automorphisms.

 We can use these maps to generate all the sheets in the full Cayley 
 tree. Suppose we have a domain with route $\{ i,j,k ,\ldots \}$
 and we apply the automorphism $A_{\alpha}$ to all  points on the 
 route.  From (\ref{amap}) this will generate a new route
 $\{\alpha, i\minus \alpha,j\plus \alpha,k \minus \alpha ,\ldots \}$.
 For instance, if we apply the map $A_{\alpha}$ to the route  
 $\{ m \}$
 from ${\cal D}$ to  ${\cal D}_m$, we obtain the route
 $\{ \alpha,  m \minus \alpha \}$ to the domain 
 $D_{ \alpha,  m \minus \alpha }$. Thus the map that takes $\cal D$
 to ${\cal D}_{ij}$ is $A_i A_{i+j}$.

 Iterating, we find that the map that takes $\cal D$
 to ${\cal D}_{ijk \ldots mn}$ is
 \be \label{map1}
 A_i A_{i+j} A_{j+k}  \cdots A_{m+n} \period \ee
 We must have
 \be \label{AB}
 A_i^2 \eq 1 \comma \ee
 since applying the same map twice merely returns $p$ to  
 the previous domain. 

 Let us refer to the general Riemann surface we have just described
 as ${\cal G}$. It consists of infinitely many Riemann sheets, 
 each sheet corresponding to a site on a Cayley tree, 
 adjacent sheets corresponding to adjacent points on the tree.
 A Cayley tree is a huge graph: it contains no circuits and is 
 infinitely dimensional, needing infinitely many integers 
 to specify all its sites.

 Any given function will have a  Riemann surface that can be 
 obtained  from ${\cal G}$ by identifying
 certain sites with one another, thereby creating circuits
 and usually reducing the graph to one of finite dimensionality. 

 {From} (\ref{amap}), the maps $A_0, A_1,\ldots , A_{N-1}$ leave 
 $t_p$ unchanged.
 We shall often find it helpful to regard $t_p$ as a fixed
 complex number, the same in all domains,  and to consider the 
 corresponding values of 
 $x_p, y_p$ (and the hyperelliptic variables $z_p, w_p$) in 
 the  various domains. To within factors of $\omega$, the
 variables  $x_p$ and $y_p$
 will be the same as those for $\cal D$ in even domains, while
 they will be interchanged on odd domains.

 \begin{center}
  ({\it \large  a} ) {\it \large  Analytic continuation of  
    $\tau_2(p,q)$}
 \end{center}

 Consider  $\tau_2(p,q)$ as a function of $q$ (so replace
 $p$ by $q$ in the previous discussion of the domains). More 
 specifically,
 think of it as a function of the complex variable $t_q$. For 
 $q \in {\cal D}$ (and  $p \in {\cal D}$), it is apparent 
 from (\ref{deftau2}) that
 $\tau_2(p,q)$ is an analytic function of $t_q$ except for the 
 {\em single}
 branch cut $ {\cal B}_{N-1}$, being single-valued across the other
 $N-1$ cuts. Let $q' = A_i \, q$, so $q' \in {\cal D}_i$, and define
 \ba \delta (i,j) =  &  \! \!  1 &  \; \; \; {\rm if \; \, } 
 i = j , \;  \;  {\rm mod}  \;  N , \nonumber \\
   \delta (i,j) =  & \! \!    0 & \; \; \; {\rm else } \, . \ea
 Then it follows from (\ref{prodtau}) that
 \be \label{defAi}
 \tau_2(p,q') \eq v_{pq}^{-\delta (i,N-1)} \,  \tau_2(p,q) \ee
 for $i = 0, 1, \ldots, N-1$, where
 \be \label{defvpq} v_{pq} \eq \frac{\alpha(p,q)}{\alpha(p,q')} \eq 
    \frac{(x_p^N-y_q^N) (y_p^N-x_q^N)}
      {(x_p^N-x_q^N) (y_p^N-y_q^N)} \eq 1/v_{pq'}  \ee
 and $x_{q'}^N = y_q^N, y_{q'}^N = x_q^N$. 


 Eqn. (\ref{defAi}) defines the mapping $A_i$ applied to the function 
 $\tau_2(p,q)$ of $q$. Iterating,
 it follows that if  
 $q'' = A_i\,  A_{j} \, A_{k} \, q$, then
 \be \label{acq}
      \tau_2(p,q'') \eq v_{pq}^{- m } \, \tau_2(p,q) \comma \ee
 where $m =  \delta (i,N-1) - \delta(j,N-1) + \delta (k,N-1) $.

 We can also keep $q$ fixed in $\cal D$ and consider the analytic
 continuation of $\tau_2(p,q)$ as $p$ moves from sheet to sheet. 
 If $p' = A_i p$, so $p \in \cal D$, $p' \in {\cal D}_i$, we can 
 verify from (\ref{deftau2}) that
 \be \label{acp}
    \tau_2(p',q) \eq (\omega^{-i} t_p-\omega t_q)^2  /\tau_2(p,q) 
     \sep t_{p'} = t_p \period \ee
 Also, from (\ref{defalpha}), 
 \be \label{alpha2} \alpha(p,q) \alpha(p',q) \eq 
 (t_p^N-t_q^N)^2 \period \ee


 Eqn. (\ref{acp}) defines the $A_i$ for the function  $\tau_2(p,q)$ of
 $p$. Iterating, it follows that if  
 $p'' = A_i\,  A_{j} \, A_{k} \, p$, then
 \be
 \tau_2(p'',q) \eq (\omega^{-i} t_p-\omega t_q)^2 \, (\omega^{-j} 
   t_p-\omega t_q)^{-2}\, (\omega^{-k} t_p-
  \omega t_q)^2  / \tau_2(p,q)  \period \ee

 Note that for both $\tau_2(p,q) \rightarrow $ 
 $\tau_2(p,q'') $ and   $\tau_2(p,q) \rightarrow $ $\tau_2(p'',q) $, 
 it is true that
 $A_i A_j A_k \eq A_k A_j A_i $.





 \begin{center}
 \section*{4. Hyperelliptic parametrization for $N=3$}
 \end{center}

 \setcounter{equation}{0}
 \renewcommand{\theequation}{4.\arabic{equation}}

 Hereinafter we restrict our attention to the case $N = 3$. A 
 parametrization of $\{ x_p,y_p,\mu_p,t_p \}$ was developed in 
 previous papers, in terms of a  ``nome'' $x$ and two 
 related parameters $z_p, w_p$ (\Baxter 1991b, 1993a, 1993b, 1998).
 The nome $x$ is like $k$ and $k'$ in that it is a constant: it is 
 not to be confused with the rapidity variable $x_p$.

 However, in (\Baxter 2006)
 we showed that the function ${\cal S}_p$ could not
 be expressed as a single-valued function of these original 
 parameters $z_p,w_p$. This is 
 because $z_p,w_p$ have the same values (for given $t_p$) in the 
 domains ${\cal D}_{021}$ and  ${\cal D}_{211}$, whereas ${\cal S}_p$
 has different values therein. These domains are obtained by the maps
 $A_0 A_2 A_0$, $A_2 A_0 A_2$, respectively, and indeed we see from
 (\ref{acq}) and (\ref{acp}) that these maps give different 
 results for $\tau_2(p,q)$ for both the $q$ and the $p$
 variables. Thus $\tau_2(p,q)$ cannot be a single-valued function,
 either of   $z_q, w_q$ for fixed $p$, or of  $z_p, w_p$ for fixed 
 $q$.

 The same problem occurs with the domains ${\cal D}_{110}$, 
 ${\cal D}_{220}$ and the corresponding maps $A_1 A_2 A_1$, 
 $A_2 A_1 A_2$.

 The situation is not lost. We also showed in (\Baxter 2006) that
 there is another way of parametrizing $k, x_p, y_p, \mu_p, t_p$. 
 This alternative way preserves the property that the nome $x$ is 
 small 
 at low temperatures ($k'$ small). It can be obtained from the 
 original  parametrization  by leaving $x, z_p, w_p$ unchanged
 and transforming $k, k', x_p, y_p,\mu_p,t_p $ according to the rule:
 \be \label{kmap}
 k, k', x_p,y_p,\mu_p,t_p \rightarrow k^{-1}, ik'/k, 1/x_p, y_p,
 \omega^{-1/4} x_p \mu_p, y_p/x_p  \ee
 (taking $ \omega^{-1/4} = \e^{-\i \pi/2 N}$). This mapping leaves the 
 relations (\ref{prlns}), (\ref{kkp}) unchanged.

 Doing this, eqn. (21)  of (\Baxter 1993a) becomes
 \be \label{defx}
 -{k'}^2 = 27 x \prod_{n =1}^{\infty} 
 \left( \frac{1-x^{3n}}{1-x^n} \right)^{12} \comma \ee
 while the two equivalent relations  (4.5), (4.6) of (\Baxter 1993b)
 remain unchanged: 
 \bd \label{eq4.5}
 w = \prod_{n=1}^{\infty} \frac{(1-x^{2n-1} z/w) (1-x^{2n-1} w/z)
 (1-x^{6n-5} zw) (1-x^{6n-1} z^{-1} w^{-1})} 
 {(1-x^{2n-2} z/w) (1-x^{2n} w/z)
 (1-x^{6n-2} zw) (1-x^{6n-4} z^{-1} w^{-1})} \ed
  \be \label{eq4.6}
 \frac{z}{w} = \prod_{n=1}^{\infty} \frac{(1-x^{2n-2} /w) (1-x^{2n} w)
 (1-x^{6n-4} z^2/w) (1-x^{6n-2} w/z^2)} 
 {(1-x^{2n-1} w) (1-x^{2n-1} /w)
 (1-x^{6n-5} w/z^2) (1-x^{6n-1} z^2/w)} \period \ee

 These $z, w$ are related to $x_p, y_p, \mu_p, t_p$ by various
 elliptic-type equations that we shall give below, being the 
 arguments (more precisely the 
 exponentials of $\i$ times the arguments) of Jacobi elliptic
 functionsof nome $x$. Thus $x$, like $k, k'$, is a constant,
 while $z, w$ are two more rapidity variables, dependent on $p$. 
 We shall them as $z_p, w_p$.
 

 First we introduce  elliptic-type  functions
 \bd 
 h(z) =  1/h(z^{-1}) = \omega^2 \, h(x z) = - \omega^2 
 \prod_{n=1}^{\infty}  \frac{ (1-  \omega x^{n-1} z) 
 (1-\omega^2 x^n/z) } { (1-\omega^2 x^{n-1} z) 
 (1- \omega x^n/z) } \comma \ed
 \bd 
 \phi_b(z) \eq    \prod_{n=1}^{\infty} 
 \frac{(1-x^{3n-2} /z) \, (1-x^{3 n-1}z)}{(1-x^{3n-2} z) \, 
 (1-x^{3 n-1}/ z)} \comma \ed
  \be \label{defphi}
  \phi(z) \eq 1/\phi(z^{-1}) \eq \phi(x^3 z)  \eq  z^{1/3} \phi_b(z)
  \ee
 so that $h(1) = \phi(1) = 1$, and define two sets of parameters 
 $\zp, \zpb$ in terms of $z_p, w_p$:
 \be \label{defzp}
  \zp_1 = z_p \sep \zp_2 = -1/w_p \sep  \zp_3 = -w_p/z_p \comma \ee
 \bd \zpb_j = \zp_{j+1}/\zp_{j-1} \comma \ed
 extending $\zp_j$, $\zpb_j$ to all $j$ by 
 \be \zp_{j+3} = \zp_j \sep \zpb_{j+3} = \zpb_j 
 \period \ee
 Thus
 \be  \label{defzpb}
 \zpb_1 = z_p/w_p^2 \sep \zpb_2 = -w_p/z_p^2 \sep \zpb_3 = - z_p w_p 
  \period \ee
 We shall also need certain cube roots $\zpt_j$ of $x \, \zpb_j$, 
 choosing them so that
 \be \label{defzpt}
 \zpt_j = (x \, \zpb_j)^{1/3} \eq \omega \, \zp_{j+1}\,  \zpt_{j-1}
  \period \ee

 Then after applying the mapping (\ref{kmap}), the  equations 
 (27), (32) of (\Baxter 1993a) become
 

 \be \label{eq27}
 y_p/ x_p = - \omega^{-j} h(\zp_j)  \comma \ee
 \be  \label{eq32}
 \frac{y_p}{x_p \mu_p^2} = - \omega^{-j}  \phi ( x \zpb_j) 
  = - \omega^{-j}  \zpt_j \phi_b ( x \zpb_j)\comma 
 \ee
 for $j=1,2,3$. The second form of (\ref{eq32}) is to be preferred
 as it fixes the choice of the leading cube root factor in the 
 function $\phi(z)$, so fixing $\zpt_j$.

 Similarly, replacing $p$ by $q$ in the above equations, we define 
 $\zq_j, \zqb_j, \zqt_j$, and hence $x_q, y_q, \mu_q, t_q$, in terms 
 of two related variables $z_q, w_q$.
 \vspace{5mm}
 \begin{center}
     {\large $( a)$}  {\it \large The low-temperature limit  }
 \end{center}

 At low temperatures  $x, k'$  are small. For $p \in {\cal D}$ (and 
 $t_p$ not close to a cube root of unity)  we can choose $z_p,w_p$ to
 tend to non-zero limits as $x \rightarrow 0$. Then  (\ref{eq4.6}) 
 both give $w_p = z_p +1$. Also, $x_p \rightarrow 1$, so 
 (\ref{prlns}) and (\ref{eq27}) determine   $x_p, y_p$ uniquely 
 (with the same value for $j=1,2,3$). Then $\mu_p^3$
 can be calculated from the last of the eqns. (\ref{prlns}), and 
 $\mu_p$ itself from (\ref{eq32}). Hence 
 when $x, k'$ are small
 \bd k'^2 = -27 x \sep w_p = z_p+1 \sep x_p = 1\comma \ed
 \bd   y_p = \frac{\omega^2-z_p} {\omega-z_p} \sep
  k' \mu_p  \eq  - 3 (1-\omega^{j} y_p)  \zpt_j \comma \ed
 for $j = 1, 2, 3$. Note that $x_p, y_p, z_p, w_p$, $\zp_j, \zpb_j$ 
 are of order unity,
 but the $\zpt_j$ are of order $x^{1/3}$, while $\mu_p$ is of order
 $x^{-1/6}$.

 \begin{center}
  {\large $( b)$} {\it \large  Mappings }
 \end{center}

  The effect of the automorphisms
 $R, U, V$ on $z_p, w_p$ is
 \bd z_{Rp} = -x w_p \sep z_{Up} = -w_p \sep
 z_{Vp} = -1/ w_p \comma \ed
 \be \label{zwautos5}
 w_{Rp} = w_p/z_p \sep w_{Up} = - z_p \sep
 w_{Vp} = z_p/ w_p \period \ee
 The mappings $U, V, M$ take a point $p$ within $\cal D$ to another
 point within $\cal D$:
 \ba \label{UVMmaps}
  p' = Up: &  \zp'_j = 1/\zp_{3-j} , & \! \! \!  \zpb'_j = \zpb_{-j} 
 \, , \; \zpt'_j = \omega^{1-j} \zpt_{-j} \comma \nonumber \\
 p' = Vp: & \zp'_j = \zp_{j+1} , & \! \! \zpb'_j = \zpb_{j+1} \, , \; 
 \zpt'_j = \zpt_{j+1}\comma \nonumber \\
  p' = Mp: & \zp'_j = \zp_j \sep & \del \zpb_j' \eq  \zpb_j \sep 
 \; \zpt'_j \eq \omega \, \zpt_j\period \ea
 and if $p' = Rp$, then $\zpt'_j = \omega^{2-j} x^{1-\delta(j,2)}/
 \zpt_{j+1}$. 

 Also, if $p' = A_i p$, then 
 \bd 
 z_{p'} = x^{2-i-3 \delta(i,0)}/z_p \sep w_{p'} \eq x^{i-1}/w_p \ed
 \be \label{Aizw}
 \zp'_j \eq x^{\delta(i+j,2)-\delta(i+j,1)}/\zp_j \sep 
 ( \zpt_j) ' = \omega^{-i-j} x^{1-\delta(i+j,0)}/\zpt_j \comma \ee
 for $i = 0,1,2$. We see that in this new parametrization  
 $A_0 A_2 A_0$ does {\em not} have the same effect on $z_p, w_p$ as 
 $A_2 A_0 A_2$ (nor do $A_1 A_2 A_1$ and $A_2 A_1 A_2$), so we no 
 longer have the problem referred to at the beginning of this section. 


 \setlength{\unitlength}{1pt}
 \begin{figure}[hbt]

 \begin{picture}(420,260) (0,0)

 \put (196,105) {\circle{15}}
 \put (193,101) {0}
 \put (203,105)  {\line(1,0) {44}}
 \put (190,111)  {\line(-3,5) {22}}
 \put (190,99)  {\line(-3,-5) {22}}
 \put (163,101) {\protect \footnotesize (0,0) }
 \put (196,83) {\large ${\cal D}$}

 \put (107,155) {\circle{15}}
 \put (104,151) {1}
 \put (114,155)  {\line(1,0) {44}}
 \put (101,161)  {\line(-3,5) {22}}
 \put (101,149)  {\line(-3,-5) {22}}
 \put (74,151) {\protect \footnotesize (1,2) }

 \put (107,55) {\circle{15}}
 \put (104,51) {2}
 \put (114,55)  {\line(1,0) {44}}
 \put (101,61)  {\line(-3,5) {22}}
 \put (101,49)  {\line(-3,-5) {22}}
 \put (74,51) {\protect \footnotesize (2,1) }

 \put (287,155) {\circle{15}}
 \put (284,152) {1}
 \put (294,155)  {\line(1,0) {44}}
 \put (281,161)  {\line(-3,5) {22}}
 \put (281,149)  {\line(-3,-5) {22}}
 \put (248,152) {\protect \footnotesize (-2,-1) }

 \put (287,55) {\circle{15}}
 \put (284,52) {2}
 \put (294,55)  {\line(1,0) {44}}
 \put (281,61)  {\line(-3,5) {22}}
 \put (281,49)  {\line(-3,-5) {22}}
 \put (248,52) {\protect \footnotesize (-1,-2) }

 \put (196,5) {\circle{15}}
 \put (193,2) {1}
 \put (203,5)  {\line(1,0) {44}}
 \put (190,11)  {\line(-3,5) {22}}
 \put (160,2) {\protect \footnotesize (1,-1) }

 \put (196,205) {\circle{15}}
 \put (193,202) {2}
 \put (203,205)  {\line(1,0) {44}}
 \put (190,199)  {\line(-3,-5) {22}}
 \put (160,202) {\protect \footnotesize (-1,1) }

 \put (158,148) {\line(1,0) {15}}
 \put (158,162) {\line(1,0) {15}}
 \put (158,148) {\line(0,1) {14}}
 \put (173,148) {\line(0,1) {14}}
 \put (163,151) {2}
 \put (178,151) {\protect \footnotesize (0,1) }

 \put (158,48) {\line(1,0) {15}}
 \put (158,62) {\line(1,0) {15}}
 \put (158,48) {\line(0,1) {14}}
 \put (173,48) {\line(0,1) {14}}
 \put (163,51) {1}
 \put (178,51) {\protect \footnotesize (1,0) }

 \put (247,-2) {\line(1,0) {15}}
 \put (247,12) {\line(1,0) {15}}
 \put (247,-2) {\line(0,1) {14}}
 \put (262,-2) {\line(0,1) {14}}
 \put (252,1) {2}
 \put (267,1) {\protect \footnotesize (0,-2) }
 \put (244,18) {\large $Z$}

 \put (247,98) {\line(1,0) {15}}
 \put (247,112) {\line(1,0) {15}}
 \put (247,98) {\line(0,1) {14}}
 \put (262,98) {\line(0,1) {14}}
 \put (252,101) {0}
 \put (267,101) {\protect \footnotesize (-1,-1) }

 \put (247,198) {\line(1,0) {15}}
 \put (247,212) {\line(1,0) {15}}
 \put (247,198) {\line(0,1) {14}}
 \put (262,198) {\line(0,1) {14}}
 \put (252,201) {1}
 \put (267,201) {\protect \footnotesize (-2,0) }
 \put (244,182) {\large $Y$}

 \put (338,148) {\line(1,0) {15}}
 \put (338,162) {\line(1,0) {15}}
 \put (338,148) {\line(0,1) {14}}
 \put (353,148) {\line(0,1) {14}}
 \put (343,151) {2}
 \put (358,151) {\protect \footnotesize (-3,-2) }

 \put (338,48) {\line(1,0) {15}}
 \put (338,62) {\line(1,0) {15}}
 \put (338,48) {\line(0,1) {14}}
 \put (353,48) {\line(0,1) {14}}
 \put (343,51) {1}
 \put (358,51) {\protect \footnotesize (-2,-3) }

 \put (66,98) {\line(1,0) {15}}
 \put (66,112) {\line(1,0) {15}}
 \put (66,98) {\line(0,1) {14}}
 \put (81,98) {\line(0,1) {14}}
 \put (71,101) {0}
 \put (86,101) {\protect \footnotesize (2,2) }
 \put (49,101) {\large $X$}

 \put (66,198) {\line(1,0) {15}}
 \put (66,212) {\line(1,0) {15}}
 \put (66,198) {\line(0,1) {14}}
 \put (81,198) {\line(0,1) {14}}
 \put (71,201) {1}
 \put (86,201) {\protect \footnotesize (1,3) }

 \put (66,-2) {\line(1,0) {15}}
 \put (66,12) {\line(1,0) {15}}
 \put (66,-2) {\line(0,1) {14}}
 \put (81,-2) {\line(0,1) {14}}
 \put (71,1) {2}
 \put (86,1) {\protect \footnotesize (3,1) }

 \end{picture}

 \vspace{1.5cm}
 \caption{\footnotesize The honeycomb lattice formed by the 
 hyperelliptic variables $z, w$.
 Circles (squares) denote sites of even (odd) parity.}
 \label{honeylatt}
 \end{figure}
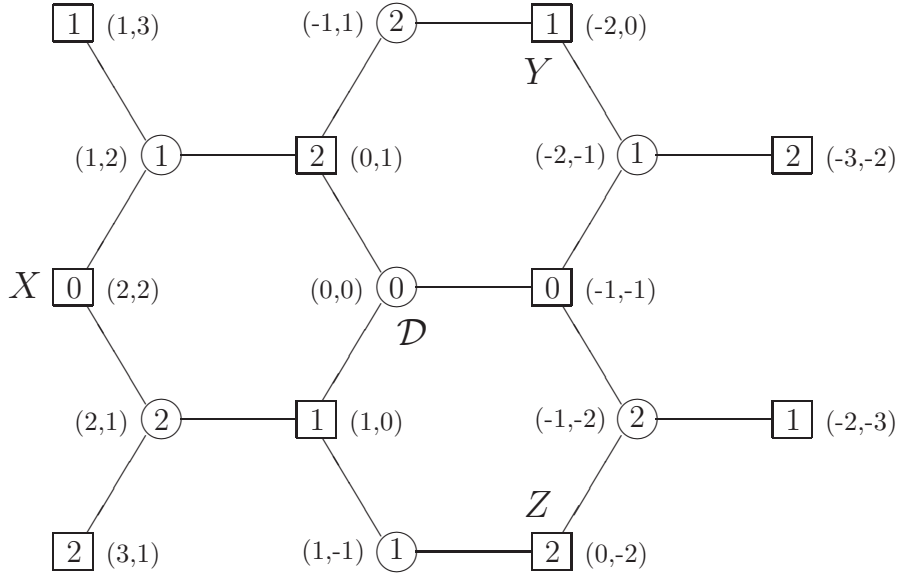

 If we identify Riemann sheets that have the same values of $z_p, w_p$
 for given $t_p$, then $\cal G$ ceases to be a Cayley tree and becomes
 the two-dimensional honeycomb lattice of Figure \ref{honeylatt}.

 To see this, note that if $z_p^0, w_p^0$ are the values of $z_p, w_p$ 
 on the central sheet 
 $\cal D$, then it follows from (\ref{Aizw}) that on any Riemann sheet 
 the analytic continuations of $z_p, w_p$ (for a given value of $t_p$) 
 are
 \be \label{anzw}
 z_p \eq x^m \, \left( z_p^0 \right)^{\pm 1} \sep w_p \eq x^n \, 
 \left( w_p^0 \right)^{\pm 1} \comma \ee
 choosing the upper (lower) signs on sheets of even (odd) parity.
 Here  $m,n$ are integers satisfying
 \ba 
 m+n & = &  0 \; \; \; ( {\rm mod} \; 3 ) \; \; {\rm on \; \;
 even \; \; sheets } \comma \nonumber \\
 & = &  1 \;  \; \; ({\rm mod} \; 3 ) \; \; {\rm on \; \;
 odd \; \; sheets } \period \ea
 The Riemann surface
 for $z_p, w_p$ therefore corresponds to a two-dimensional
 graph $\cal G$, each site of  $\cal G$ being specified by the 
 two integers $m, n$ and corresponding to a Riemann sheet of
 the surface.

 This $\cal G$ is indeed the honeycomb lattice shown in Figure
 \ref{honeylatt}. Adjacent sites correspond to 
 adjacent Riemann sheets. Sheets of even parity correspond to sites 
 represented by circles, those of odd parity  are represented by 
 squares. If $i$ is the integer shown inside the circle or square, 
 then on even sites $x_p \simeq \omega^i$, and on odd sites 
 $y_p \simeq \omega^i$. The numbers shown in brackets alongside each 
 site are the integers $m,n$ of (\ref{anzw}): we refer to the 
 corresponding sheet as ``the sheet $(m,n)$''.

 We can trace this reduction of $\cal G$ to the fact that 
 the automorphisms  $A_i$ applied to
 $z_p, w_p, \zpt_j$ satisfy 
 \be \label{Aijk}
 A_i A_j A_k \eq A_k A_j A_i {\rm \; \; for \; \; all\; \; }
    i,j,k \period \ee 
 For instance, the relation $A_2 A_1 A_0 = A_0 A_1 A_2$ implies that
 $D_{221} = D_{011}$, and indeed we see from Figure  \ref{honeylatt}
 that these are the two three-step routes from $\cal D$ to $Y$.
 Similarly for the two routes to $Z$, or the two routes to $X$.

 We noted in the previous section that the function $\tau_2(p,q)$, 
 considered as a function of either $p$ or $q$,
 also satisfies (\ref{Aijk}). It follows that its Riemann surface
 can be embedded in that of the reduced graph $\cal G$.

 We focus on the $q$ dependence of $\tau_2(p,q)$. Let its value
 when $p,q \in {\cal D}$ (given by eq. \ref{deftau2}) be 
 $\tau_2^0(p,q)$. Then by iterating (\ref{defAi}) we find that
 on the Riemann sheet $(m,n)$
 \be \tau_2(p,q) \eq v_{pq}^r \, \tau_2^0(p,q)  \comma \ee
 where 
 \ba r & = & (m-2n)/3 {\rm \; \; \; \; on  \; \; even \; \; sheets,} 
   \nonumber \\
 r & = & (m \minus 2n \minus 1)/3 {\rm \; \; on  \; \; odd \; \; 
 sheets.}  \ea

 \begin{center}
  {\large $( c)$} {\it \large   The function $T(p,q)$}
 \end{center}


 We can eliminate the distinction between even and odd sheets
 by working, not with $\tau_2(p,q)$, but instead with the closely 
 related function 
 \be \label{defT}
 T(p,q) \eq \frac{\tau_2(p,q)^3} {\alpha(p,q)}
  \eq \frac{\tau_2(p,q)^2}{\tau_2(p,Vq) \, \tau_2(p,V^2 q)} \comma \ee
 (using eqn.  \ref{prodtau}). Then (\ref{defAi}) becomes
 \be \label{TAi}
 T(p, A_i \, q) \eq v_{pq}^{1-3\delta(i,N-1)} \, T(p,q)  \ee
 for $i = 0, 1, 2$. If  $T_0(p,q)$ is the value of  $T(p,q)$ in 
 $\cal D$, then it 
 follows that for a given value of $t_q$,
 \be T(p,q) \eq v_{pq}^{m-2n} \, T_0(p,q)  \comma \ee
 on all sheets $(m,n)$.

 It follows that  $T(p,q)$ is a single-valued meromorphic function on 
 the Riemann surface $\cal G$, and that the orders of its zeros and 
 poles are linear in the integers $m$, $n$  of (\ref{anzw}) (with $p$ 
 replaced by $q$ therein). This suggests 
 that it may be possible to write
 $T(p,q)$ as a product of functions of $z_q, w_q$, and indeed we 
 shall find that this is the case. 

 We shall explicitly exhibit the dependence of 
 $T(p,q)$ on $z_q, w_q$ by writing it as $T_p(z_q,w_q)$. Then the 
 three relations (\ref{TAi}) become
 \bd
   T_p( x^{-1}/ z_q, x^{-1}/ w_q ) \eq
  T_p(x/z_q, 1/w_q ) \eq v_{pq} T_p(z_q,w_q) 
 \comma \ed
 \be \label{Trelns}
  T_p(1/z_q, x/w_q ) \eq v_{pq}^{-2} \, T_p(z_q,w_q) 
 \period \ee

 If $p$ is fixed within $\cal D$, then from (\ref{deftau2}) and 
 (\ref{defT}), $T_p(z_q,w_q)$ is bounded and has no zeros or poles for 
 $q \in {\cal D}$. (Both $\tau_2^3$ and $\alpha$ become infinite as  
  $y_q, t_q \rightarrow \infty$, but their ratio remains finite.) 
 Together with this restriction, the relations  (\ref{Trelns}) define
 $T_p(z_q,w_q)$ to within a multiplicative factor independent of $q$. 
 To see this, suppose we had two such solutions, then (\ref{Trelns})
 would imply that their ratio was unchanged by $q \rightarrow A_i q$,
 for $i = 0, 1, 2$. This would mean that the ratio, considered as a
 functions of $t_q$, did not have the  branch cuts ${\cal B}_0$, 
 ${\cal B}_1$, ${\cal B}_2$. It would therefore be an entire bounded
 function of $t_q$, and hence by Liouville's theorem a constant 
 (independent of $q$). This constant could be determined from 
 the product relation (\ref{prodtau}), which now takes the simple form
 \be   \label{Tprod}  T_p(z_q,w_q) 
 T_p(-1/w_q,z_q/w_q) T_p(-w_q/z_q,-1/z_q)   \eq 1  \period \ee
 
 We shall also use the $p$-relation (\ref{acp}). Together with 
 (\ref{alpha2}) this implies that 
 \be \label{Tpprd}
   T(p,q) T(A_ip,q) \eq \frac{(t_p - \omega^{i+1}t_q)^6}
      {(t_p^3-t_q^3)^2 } \period \ee






 \begin{center}
 \section*{5. Various $p, q$ relations}
 \end{center}

 \setcounter{equation}{0}
 \renewcommand{\theequation}{5.\arabic{equation}}

 First we present various relations that enable one to express 
 certain rational functions (including $v_{pq}$) of 
 $x_p, \ldots, t_q$ as products of elliptic functions of 
 $z_p, w_p, z_q, w_q$.

 Some can be obtained by applying the rule (\ref{kmap}) to eqn 34 of 
 (\Baxter 1993a) or to eqn 4.9 of (\Baxter 1993b). In particular, we 
 obtain
 \be \label{eq34}
 \frac{\mu_q (1-x_q y_p) }{\mu_p (1-x_p y_q) } 
  \eq \phi (z_q/z_p) \,  \phi(w_q/w_p) \period \ee

 Applying the automorphism $U$ to the $q$ variable in (\ref{eq34}), 
 and then using the $V$ and $R$ automorphisms,  we obtain the two sets 
 of relations
 \bd
 \frac{\mu_p (y_q - \omega^j x_p)}{\mu_q (y_p - \omega^i x_q)} \eq
 \omega^{j-i} \frac{\phi( x^{m} \, \zp_{m-i+1} \, \zq_{m-j-1})}
 {\phi( x^{m} \, \zp_{m-i-1} \, \zq_{m-j+1})} \comma \ed
 \be \label{req2}
 \frac{\mu_p \mu_q(x_q - \omega^j x_p)}{y_p - \omega^i y_q} \eq
 \omega^{m-i} \,   \frac{\phi( x^{j+1} \, \zp_{m-i+1} /\zq_{m-j-1})}
 {\phi( x^{j-1} \, \zp_{m-i-1} / \zq_{m-j+1})} \comma \ee
 true for all integers $i, j, m$.

 The function $\phi(z)$ has a leading factor $z^{1/3}$, which gives a 
 contribution $z_q w_q/(z_p w_p)^{1/3}$ to the RHS of (\ref{eq34}). 
 This  cube root can and should be chosen to be $\zqt_3/\zpt_3$. The 
 corresponding contributions in  (\ref{req2}) are 
 $\zpt_{m-i}/\zqt_{m-j}$ and $ \zpt_{m-i} \, \zqt_{m-j}$.

   We define the elliptic  function
 \be \label{defgr}
 g_r(z) \eq \prod_{n=1}^{\infty} (1-x^{rn-r} z) (1-x^{rn}/z) 
  \ee
 for integer  $r$ (in particular $r = 1$ and  3), satisfying
 \bd g_r(x^r z) \eq g_r (1/z) \eq -z^{-1} \, g_r(z) \period \ed
 Then $h(z) = \omega^2 g_1(\omega z)/g_1(\omega^2 z)$ and
 $\phi_b(z) = g_3(x/z)/g_3(x z)$, so (\ref{eq27}), (\ref{eq32})
 can be used to relate various $g_1$ and $g_3$ functions. In 
 particular, from (\ref{defzpt}) and (\ref{eq32}),
 \be \label{eq32corr}
 g_3(\zpb_j)/g_3(\zpb_i) \eq \omega^2 \zp_k^2 \, 
  g_3(x^2 \zpb_j)/g_3(x^2 \zpb_i) \comma \ee
 for any cyclic permutation $\{ i, j, k \}$ of $\{ 1, 2, 3 \}$.
 We note that (\ref{defx}) can be written
 \be \label{defxa} -{k'}^2  \eq 27 x/g_3(x)^{12}  \ee
 and  set
 \bd \gamma \eq (- 27 x k^4/{k'}^2)^{1/6}  \eq
  - 3 \omega k^{2/3}/g_1(\omega)^2 \ed
 \be \label{defgamma}
  \eq k^{2/3} g_3(x)^2 \eq 1 + 7x + 8 x^2 + 
 22 x^3 + 42 x^4 +  \cdots  \period \ee
 It follows from (\ref{defgr}) and (\ref{defxa}) that these various 
 expressions for $\gamma$ are all equal.

 We also define
 \bd v_1(p,q) \eq  \frac{(y_p^3-x_q^3)(y_q^3-1/y_p^3)}{(y_p^3-y_q^3)
 (x_q^3-1/y_p^3)}  \comma \ed
 \be v_3(p,q) \eq \frac{(x_p^3-y_q^3)(x_q^3-1/y_p^3)}{(x_p^3-x_q^3)
 (y_q^3-1/y_p^3)} \comma \ee
 and note  that 
 \be \label{vvv}
 v_{pq} \eq v_1(p,q) v_3(p,q) \period \ee

 We find the following six identities and indicate  below the method 
 of their proof. The first two can also be derived from (\ref{req2})
 by applying the duality/conjugate modulus mapping of 
 the Appendix.
 
 We have also checked all the identities of this section numerically,
 for arbitrarily chosen $x, z_p, z_q$, to 25 digits of accuracy.
\bd
 \frac{y_q  - \omega^j x_p \mu_p \mu_q}
 {y_p  - \omega^i x_q \mu_p \mu_q}
 \eq \omega^{j-i} \, \frac{h(\omega^{m}  
 \zpt_{m-i-1} \zqt_{m-j+1})}
 {h(\omega^{m}    \zpt_{m-i+1} \zqt_{m-j-1})} 
 \comma \ed
 
 \be \label{newids}
 \frac{x_q \mu_q - \omega^j x_p \mu_p}{y_p \mu_q - \omega^i y_q \mu_p}
 \eq \omega^{m-i} \, \frac{h(\omega^{j-1} \, 
 \zpt_{m-i-1}/\zqt_{m-j+1})}
 {h(\omega^{j+1} \, \zpt_{m-i+1}/\zqt_{m-j-1})} 
 \comma \ee

 \be  \label{id1} 
 v_1(p,q) \eq  \prod_{j=1}^3 \frac{g_1(z_q\zp_j)}{g_1(z_q/ \zp_j) } 
        \comma \ee
 \be \label{id3}  v_3(p,q) \eq \frac{z_q^2 \, w_q^2}{x} 
   \prod_{j=1}^3  \frac{g_3(-x^2 \, z_q w_q \zpb_j) } 
   {g_3(-z_q w_q/\zpb_j)} \comma \ee

 \bd 
   \frac{\omega^2 g_1(\zp_i\, \zq_j)\,  g_1(\zq_j/\zp_i) }
 { g_1(\omega^2 \zp_i) 
 g_1(\omega^2/ \zp_i) g_1(\omega^2 \zq_j) g_1(\omega \zq_j) } \eq
 \frac{g_3(\zqb_{j}/ \zpb_{i})\,  g_3(x^2 \zpb_{i} \, \zqb_{j}) }
 { g_3(1/\zpb_{i}) 
 g_3(x/ \zpb_{i})  g_3( \zqb_{j}) g_3(x^2  \zqb_{j}) } \ed
 \be \label{eq50}
  \eq
 \frac{ \gamma \, (\omega^i t_p - \omega^j t_q)
  (\omega^{i+j} t_p t_q - 1)}
    {3 \, \omega^{i+j} t_p t_q} \comma \ee
 for $i, j, m = 1,2,3 $. 

 Letting $q \rightarrow Vq$, we see that (\ref{id1}) remains true
 if $z_q$ therein is replaced by $-1/w_q$ or by $-w_q/z_q$.
 Also,  (\ref{id3}) is true if $-z_q w_q$ is replaced by 
 $z_q/w_q^2$ or by $-w_q/z_q^2$.


 \vspace{5mm}
 \begin{center}
  {\large $( a)$} {\it \large  Proof of the identities }
 \end{center}

  We have proved the identities (\ref{newids}) - 
 (\ref{eq50}). Let $E_q$ be the ratio of the LHS to the RHS of any
 of these identities. 

 First note that {\em a priori} we expect  $E_q$ to 
 be a single-valued function of $t_q$ only if we 
 introduce the branch cuts ${\cal B}_i$ of Fig. \ref{brcuts} into
 the complex $t_q$-plane. 

 Indeed, the identities (\ref{newids})
 involve $\mu_q$ and $\zqt_j$, and these  will need 
 additional cuts linking  the ${\cal B}_i$ in order to completely fix 
 the choice of the cube root in (\ref{defzpt}). 

 The map that takes $q$ from one such choice to another is the map
 $q \rightarrow Mq$. However,  from 
 (\ref{autos}) and (\ref{UVMmaps}), this merely 
 increases (decreases) $i,j,m$ by one in the first (second) set of
 the identities (\ref{newids}). If we take some symmetric function
 $\Phi_q$ of all the $E_q$ (for all $i,j,m$), then  $\Phi_q$
 will be invariant under $q \rightarrow Mq$, which means that
 these extra cuts are not needed for this function.

 The next step is to use (\ref{amap}) and 
 (\ref{Aizw}) to show that the mapping $q \rightarrow A_i q$
  merely permutes, and possibly inverts, the $E_q$ of each set of 
 identities. This is true for each of the sets of identities
 (\ref{id1} ) - (\ref{eq50}), but for (\ref{newids}) the two sets are 
 interchanged by this mapping.

 Again, let  $\Phi_q$ be some symmetric function of all the $E_q$
 (and if necessary their inverses) within a set (e.g. the sum of 
 the fifth powers of every   $E_q$ in  one of the eqns. \ref{eq50}, 
 summed over $i$ and $j$), now regarding the two identities  
 (\ref{newids}) as forming one combined
 set. Then $\Phi_q$ will be unchanged by each of the three mappings
 $q \rightarrow A_i q$, for $i = 1,2,3$. This means that it has the
 same value on either side of any of the branch cuts  ${\cal B}_i$.
 The cuts are therefore unnecessary: $\Phi_q$ is a {\em single-valued
 function} of the complex variable $t_q$.


 The only possible singularities of $\Phi_q$ are therefore poles, 
 arising from poles, and possibly zeros, in the $E_q$.
 The only places these can occur are when $t_q^3 = t_p^3$, 
 $t_q^3 = 1/t_p^3$, $t_q = 0$ and $t_q = \infty$. We can restrict 
 our attention to $p, q \in {\cal D}$, since $\Phi_q$ is unchanged (for 
 a given value of $t_q$) by crossing the ${\cal B}_i$. Thus $|\mu_q|$
 and $|\mu_p|$ are both greater than one, and $x_q, x_p$ each 
 lie in a region near the points $x_q =1$, $x_p = 1$. This means that
 the only points to consider are:
  \ba \label{zeros}
   1)   \{ x_q,y_q,\mu_q,\zq_j, \zqb_j, \zqt_j \} & = & 
  \{ x_q,\omega^i y_q , \omega^m \mu_q , \zq_{j+i} , \zqb_{j+i}, 
   \omega^m \zqt_{j+i} \} \comma \nonumber \\
   2)   \{ x_q,y_q,\mu_q,\zq_j, \zqb_j, \zqt_j \}  & = &
  \{ 1/x_q,\omega^i /y_q , \omega^{m-1/2} x_q\mu_q/y_q , \nonumber \\
   && \;\;\;\;\;  \zq_{-j-i}^{-1} , \zqb_{-j-i}, 
   \omega^{m-j-1} \zqt_{-j-i} \} \comma  \\
  3)   \{ x_q,y_q,\mu_q,\zq_j, \zqb_j, \zqt_j \} & = & 
  \{ k^{1/3} , 0 , \omega^m /{k'}^{1/3} , \omega^2 , 1, 
   \omega^m x^{1/3} \} \comma \nonumber \\
  4)  \{ x_q,y_q,\mu_q,\zq_j, \zqb_j, \zqt_j \} & = & 
  \{ k^{1/3} , \infty , \omega^m (-k/k')^{1/3} y_q , \omega , 1, 
   \omega^{m-j} x^{1/3} \} \comma \nonumber \ea
  for all $i$ and $m$.

  Each $E_q$ can be written as a ratio of pole-free functions. In 
 every case, if the numerator (denominator) has a zero at one of the 
 points (\ref{zeros}), then so does the denominator (numerator), and 
 both zeros are simple. Thus no $E_q$ has a pole or zero at any of 
 the above points.

   The function $\Phi_q$ is therefore a single-valued  and analytic 
 function in the complete $t_q$ plane, including the point at infinity 
 (this is the last of the points listed above). By Liouville's theorem 
 it is therefore a constant. 

 One can write down a polynomial of finite degree whose roots are the 
 $E_q$ and whose coefficients are symmetric functions $\Phi_q$. Since 
 each such coefficient is a constant, so are the roots. Thus every 
 $E_q$ is a constant. By looking at special values of $q$, e.g. one 
 of the values above, in each case we can show that the constant is 
 unity, and hence prove the identities (\ref{newids})  - (\ref{eq50}). 
 In the last two identities one needs to take the limit as 
 $y_q \rightarrow 0$ or $\infty$,  using  (\ref{defx}), 
 (\ref{eq27}), (\ref{eq32}) and (\ref{eq32corr}) - (\ref{defgamma}). 
 
 (One can streamline the procedure: for instance, each of the 
 27 identities in the first set of relations (\ref{newids}) can be 
 obtained from one of them by using the mappings $q \rightarrow Vq$,  
 $q \rightarrow Mq$, $p \rightarrow Vp$,  $p \rightarrow Mp$, so it 
 sufficient to do the last step for just one of the equations, say 
 $i=j=m=0$.)






 \begin{center}
 \section*{6. Calculation of $T(p,q)$ }
 \end{center}

 \setcounter{equation}{0}
 \renewcommand{\theequation}{6.\arabic{equation}}

 We now look for solutions of (\ref{Trelns}), using the identities 
 (\ref{vvv}) - (\ref{id3}). This leads us to define, similarly to 
 section 6 of (\Baxter 2006), the function
 \be \label{defFr}
 F_r(z)  \eq \prod_{j=1}^{\infty} \frac{(1-x^{rj} z)^j}
 {(1-x^{rj} z^{-1}  )^j}  \period \ee
 It satisfies
 \be F_r(z) \eq 1/F_r(z^{-1} ) \eq -z^{-1} g_r(z) F_r(x^r z)  
   \ee
	and is a natural extension of the elliptic function $g_r(z)$.
 
 We further define the functions
 \bd
 R_1(z) \eq \prod_{j=1}^3 F_1(z/\zp_j)/F_1(z \zp_j) \comma \ed
 \bd 
 R_3(v) \eq \prod_{j=1}^3 F_3(v/\zpb_j) F_3(x v^{-1} /\zpb_j) 
 \comma \ed
 suppressing their dependence on $p$. They have been 
 constructed so that 
 \bd R_1(x^{-m}/ \zq_i) = v_1(p,q)^m \,  R_1(\zq_i) \comma \ed
 \be \label{reln3}
 R_3(x^{1-3m}/\zqb_i) = (-\zqb_i)^m \, x^{m(3m-1)/2} \, v_3(p,q)^m  
    R_3(\zqb_i)   \comma \ee
 for $i = 1,2,3$ and  all integers $m$.  Because they are products 
 over $\zp_1,\zp_2,\zp_3$
 (or $\zpb_1$, $\zpb_2$, $\zpb_3$), they, like $v_1(p,q)$, $v_3(p,q)$, 
 $T(p,q)$, are unchanged by $p \rightarrow Vp$. They only have zeros 
 or poles when $v_1(p,q), v_3(p,q)$ have zeros or poles, i.e.
 when $x_q^3$ or $y_q^3$ equals  $x_p^{\pm 3}$ or $y_p^{\pm 3}$. None 
 of these zeros or poles occur when $p,q \in {\cal D}$.

 The factors $(-\zqb_i)^m$, $ x^{m(3m-1)/2}$ in (\ref{reln3}) are 
 irritating as they do not occur in (\ref{Trelns}). However, they are
 independent of $p$, so we may hope to remove them by introducing some 
 additional simple factor $\chi_q$ that is also 
 independent of $p$. Also, (\ref{Trelns}) is unchanged by multiplying
 $T(p,q)$ by any function $\eta_p$ of $p$ only.

 We therefore  try the ansatz
 \be  \label{ansatz}
  T_p(z_q,w_q) \eq  \eta_p \, \chi_q \, \prod_{i=1}^3 R_1(\zq_i)^{a_i}  
    R_3(\zqb_i)^{b_i} \comma \ee
 where $a_1, a_2, a_3, b_1, b_2, b_3$ are integers, to be determined.
 We first seek to satisfy the relations (\ref{Trelns}), and find that
 we can match the powers of 
 $v_{pq} = v_1(p,q) v_3(p,q)$ therein by taking
 \be a_2 = a_1 - 1 \sep a_3 = a_1+1 \sep b_2 = b_3 = 1 \sep b_1 = -2 
   \period \ee


  At this stage we are free to choose $a_1$, which corresponds to 
 multiplying $T_p(z_q,w_q)$ by a factor $R_1(\zq_1) R_1(\zq_2) 
 R_1(\zq_3)$ $= R_1(z_q) R_1(-1/w_q) R_1(-w_q/z_q)$.
 This factor has no effect on the relations  (\ref{Trelns}) and
 is bounded, with no zeros or poles, for $p,q \in {\cal D}$. From the 
 argument after  (\ref{Trelns}), it must therefore be independent of 
 $q$. Taking  $t_q = \infty$ and 
 $\zq_i = \omega$, or $t_q = 0$ and $\zq_i = \omega^2$,  we obtain the 
 identity
 \be \prod_{j=1}^3 R_1(\zq_j)  \eq R_1(\omega)^3 \eq R_1(\omega^2)^3 
 \ee
 for all $q$.

  Without loss of generality we can therefore choose $a_1=0$, 
 giving $a_2 = -1, a_3 = 1$. Substituting the ansatz (\ref{ansatz}) 
 into (\ref{Tprod}), the $R_1$ and $R_3$ functions cancel out, 
 leaving
 \be \label{chiprod}
 \eta_p^3 \, \chi_q \, \chi_{Vq} \,  \chi_{V^2q} \eq 1 \period \ee
 Since $p$ and $q$ are independent variables, $\eta_p$ must be a 
 constant (independent of both $p$ and $q$). We can absorb this 
 constant into the factor $\chi_q$ in (\ref{ansatz}), so we can 
 set 
 \be \label{etap} \eta_p = 1 \period \ee

 We can calculate $\chi_q$ from (\ref{Tpprd}). Take $i=1$ 
 therein, so  $p' = A_1 p$ and 
 $p'_j = x^{2-j}/\zp_j$ and $\zpb'_j = x^{1-3\delta(j,2)}/\zpb_j$ for  
 $j = 1, 2, 3$. Let $R'_m(z)$ be the 
 function $R_m(z)$ defined above, but with $p$ replaced by $p'$. From 
 the above definitions and properties we can verify that  
 \bd R'_1(z) R_1(z) \eq \frac{\zp_1 \zp_3 \, g_1(z \zp_1) g_1(z/\zp_1)}
 {g_1(z \zp_3) g_1(z/\zp_3) } \ed
 and
 \be R'_3(v) R_3(v) \eq  \frac{1 }{g_3(v/\zpb_2) 
   g_3(x^2 v \zpb_2)} \period \ee
 Using these, we find (after some work) from 
 (\ref{eq50})  and (\ref{ansatz}) that  (\ref{Tpprd}) is satisfied 
 for $i=1$ iff
 \be \label{chisq} \chi_q^2 \eq \frac{g_3(\zqb_2) g_3(x^2 
 \zqb_2) g_3(\zqb_3) 
  g_3(x^2 \zqb_3)}{g_3(\zqb_1)^2 g_3(x^2 \zqb_1)^2} \period \ee
 Taking $i = 2 $ or $i=3$ in (\ref{Tpprd}) merely permutes 
 $\zp_1,\zp_2,\zp_3$ in the working, which leaves the $p$-independent 
 result (\ref{chisq}) unchanged.

 We can use (\ref{eq32}) to eliminate the factors $g_3(\zqb_j)$ in 
 favour of $g_3(x^2 \zqb_j)$. Using also (\ref{chiprod}), we obtain
 \be  \chi_q \eq \frac{ g_3(x^2 \zqb_2) g_3(x^2 \zqb_3) }
  {\zqb_1 \, g_3(x^2 \zqb_1)^2} \period \ee

 Writing $\chi_q$ as $\chi(z_q,w_q)$ and using  (\ref{eq32}), we find
 that
 \bd -x \, \zqb_3 \chi(x^{-1} /z_q,x^{-1} /w_q) \eq -x \, \zqb_2 
 \chi(x /z_q,1 /w_q)  \ed
 \be \label{chirelns}
  \eq x^{-2} \, \zqb_1^{\raisebox{.4ex}{$\scriptstyle{-2}$} } \,
 \chi(1/z_q,x/w_q) \eq \chi(z_q,w_q) \period \ee

 Our ansatz (\ref{ansatz}) is now
 \be \label{finansatz} T_p(z_q,w_q) \eq \chi_q \frac{R_1(\zq_3) 
   R_3(\zqb_2) R_3 (\zqb_3) }
 { R_1(\zq_2) R_3(\zqb_1)^2 } \period \ee
 Using (\ref{reln3}) and (\ref{chirelns}), we find that this 
 expression does indeed satisfy the relations (\ref{Trelns}), so from 
 the argument following (\ref{Trelns}) we see that the expression 
 (\ref{finansatz}) 
 must be correct to within  a factor that is independent of $q$. It 
 also satisfies (\ref{Tprod}), so this factor must be a cube root of 
 unity. Since (\ref{Tpprd}) is satisfied, this root must be unity 
 itself. This completes the proof of (\ref{finansatz}).

 \vspace{5mm}

 \begin{center}
  {\large $( a)$} {\it \large Relation to the order parameter 
    function ${\cal S}(t_q)$  }
 \end{center}

 In an earlier paper (\Baxter 2006), the author considered the function
 ${\cal S}(t_q)$ that occurs in the derivation of the order parameter 
 of the chiral Potts model. This is given by  
 \be \label{defS}
 \log S(t_q ) \eq - \frac{2}{N^2}\log k + \frac{1}{2 N \pi} 
 \int_{0}^{2 \pi}  \frac{k' \e^{\i \theta} }{1-  k'
 \e^{\i \theta} }
 \;  \log\left[  \Delta (\theta )  -  t_q \right] \, 
 {\rm d} \theta \comma \ee
 $\Delta(\theta)$ being defined by (\ref{defDelta}). This is very 
 similar to the function $\tau_2(p,q)$ of this paper, in fact if $q_3$ 
 is the point (3) of eqn. \ (\ref{zeros}), with $\mu_q^N = 1/k'$ 
 and $t_q = 0$, and $q_4$ is the point (4), 
 with $\mu_q = t_q = \infty$,  then
 \be \label{Sfn}
 S(t_q)^{2N} \eq \frac{\tau_2(p_3,q)}{{k}^{4/N}  \tau_2(p_4,q)} 
 \period \ee

 For $N=3$, we can use our results (\ref{defT}), (\ref{finansatz}) to 
 express the 
 RHS of (\ref{Sfn}) in terms of our elliptic-type functions. Because
 $\zqb_j=1$ for both points (3) and (4), the $R_3$ factors cancel.
 The $R_1$ factors reinforce, giving
 \be S(t_q)^{3} \eq \frac{x_q G((-1/w_q)}{k^{1/3} G(z_q) } \comma \ee
 where
 \bd  G(z) = F_1(\omega z)/F_1(\omega^2 z) \period \ed
 This is indeed the result given in equations (57), (62) of 
 (\Baxter 2006) (with $p$ therein replaced by $q$). Because the $R_3$
 functions have cancelled, it contains no elliptic-type functions
 with arguments $z_q/w_q^2$, $-w_q/z_q^2$ or $-z_q w_q$.






 \begin{center}
 \section*{7. Summary}
 \end{center}

 \setcounter{equation}{0}
 \renewcommand{\theequation}{7.\arabic{equation}}

 For $N=3$, we have written the integral expression (\ref{deftau2}) as 
 a product of generalized elliptic functions, with arguments that are 
 ratios and  products of the variables $\zp_j, \zq_j, \zpb_j, \zqb_j$ 
 defined in (\ref{defzp}), (\ref{defzpb}). These are the variables of 
 the alternative hyperelliptic parametrization of the chiral Potts
 model.

    Is this progress? To the author it seems that the answer is indeed
 yes: functions such as the $F_r(z)$ of (\ref{defFr}) occur naturally 
 in the free energy of other models, such as the Ising, six and 
 eight-vertex models, so it is interesting to see them occurring 
 again in the three-state chiral Potts model. Indeed, it may indicate
 some intriguing relations between such models. Certainly it provides
 an explicit formulation of the meromorphic structure of $\tau_2(p,q)$
 on its Riemann surface.

   Unfortunately it is not clear how one could proceed to $N >3$.
 There seems then to be no reason to expect to be able to express 
 quantities such 
 as $v_{pq}$ as products of single-argument Jacobi elliptic functions.
 The best one can hope for is to write them as ratios of hyperelliptic
 theta functions, as in (\Baxter 1991b). These are entire functions of 
 two or more related variables, all of whose zeros are simple. To 
 write $\tau_2(p,q)$ in such a parametrization, one would need 
 to generalize the hyperelliptic theta functions to make the order of 
 the zeros increase linearly with the distance of the zero from some 
 origin. The author knows of no such generalization. Indeed, while 
 Jacobi's triple product identity enables one to write Jacobi theta
 functions as either products or sums, there seems no reason to expect
 the same of functions such as the numerators or denominators in 
 (\ref{defFr}).

   There is also a problem with extending our working to the free 
 energy of the full $N=3$ chiral Potts, which is given by the double 
 integral in eq. 46 of (\Baxter 1991a) and eq. 61 of (\Baxter 2003a). 
 The author 
 showed in (\Baxter 2003b) that the graph of the Riemann surface of 
 this function has one more dimension than $\tau_2(p,q)$. If we fix 
 $p$
 and consider the free energy as a function of $q$, then  we do not 
 have the relation (\ref{Aijk}) and we need a three-dimensional 
 lattice to represent the Riemann surface, rather than the honeycomb 
 lattice of Figure \ref{honeylatt}. This implies the need for one 
 more
 ``hyperelliptic'' variable, in addition to $z_q$ and $w_q$, and it
 quite unclear what this may be. (No such extra variable is needed 
 for the $N=2$ Ising case.)





\vspace{5mm}
\begin{center}

{\section*{Appendix A: The duality map}}
\end{center}
\vspace{-6mm}

 \setcounter{equation}{0}
 \renewcommand{\theequation}{A\arabic{equation}}

    The mappings $R, U, V, M$ leave $k, k', x$ unchanged. There is 
 another map that takes $k'$ to $1/k'$ and $p$ to $p'$, where
 \be \label{dual}
  \{ x_{p'}, y_{p'}, \mu_{p'} , t_{p'} \} \eq \{ \omega^{1/4} 
  x_{p} \mu_p,  \omega^{1/4} y_{p} /\mu_p, 1/\mu_{p} , 
  \omega^{1/2} t_p  \} \period \ee

 Let $x = \e^{-\pi \lambda} \sep z \eq \e^{\i \pi \alpha} $ and write
 the RHS of (\ref{defx}) as $r(\lambda)$. Also, write $h(z), \phi(z)$ 
 as  $h(\alpha, \lambda), \phi(\alpha, \lambda)$. Set
 \be \lambda' = 4 \pi/(3 \lambda)   \period \ee

 Then, as in \S 15.7 of (\Baxter 1980), one can establish the 
 conjugate modulus relations  
 \be \label{conjmod}
 h(\alpha, \lambda) \eq 
 \phi(- 2 \i \alpha/\lambda, \lambda') \sep
  \phi(\alpha, \lambda) \eq h(2 \i \alpha/3 \lambda , \lambda') 
  \period \ee
 Also,  $ r(\lambda) = 1/r(\lambda') $, so
 replacing $\lambda$ by $\lambda'$ does indeed invert $k'$, as 
 given by  (\ref{defx}). Further, if $\zp_j = \exp(\i \pi \alpha_j) $, 
 then we can choose
 \bd \alpha_1 + \alpha_2 +\alpha_3  = -2 \period \ed
 (This enables one to take $\alpha_j = -2/3$, for all $j$, for the
 interesting  case when $y_p=0$ and $\zp_j = \omega^{-1}$.)

 If we set
 \be \alpha'_j \eq \frac{2i}{3 \lambda} (\alpha_{j+1}-\alpha_{j-1}+ 
   i \lambda ) \comma \ee
 then  it is also true that $\alpha'_1 + \alpha'_2 +\alpha'_3  = -2$.
 In fact the mapping $\lambda, \alpha_j \rightarrow \lambda', 
 \alpha'_j$ is self-reciprocal. Further, using (\ref{conjmod}), we 
 find
 \bd h(\alpha_j, \lambda) = \phi(\alpha'_{j+1} \minus \alpha'_{j-1} 
  \! + i \lambda',\lambda')  \sep
  \phi(\alpha_{j+1} \minus \alpha_{j-1} \! + 
   i \lambda, \lambda) = h(\alpha'_j, \lambda') \period \ed

 The left-hand sides of these last two equations are the functions 
 $h(\zp_j)$, $\phi(x \zpb_j)$ of eqns.\ (\ref{eq27}), (\ref{eq32}).
 It follows that these two equations are interchanged by the 
 duality mapping (\ref{dual}), provided we also use the conjugate 
 modulus form (\ref{conjmod}) of the elliptic functions and then
 make the transformation $\lambda, \alpha_j 
 \leftrightarrow  \lambda',  \alpha'_j$.

 \begin{center}
 \subsection*{References}
 \end{center}

 \begin{description}


 \item[] Au-Yang,~H., McCoy,~B.~M., Perk,~J.~H.~H., Tang,~S. 
 \& Yan,~M.-L. 1987  Commuting transfer matrices in the chiral Potts 
 models: solutions of star-triangle equations with genus $>$ 1.
 {\it Phys. Lett. A} {\bf 123}, 219--223.

  \item[]  Baxter,~R.~J. 1980
 Hard hexagons: exact solution. 
 {\it J. Phys. A} {\bf 13}, L61--L70.

 \item[]  Baxter, R. J. 1982a
 {\it Exactly Solved Models in Statistical Mechanics},  
 London,  Academic.

 \item[]  Baxter,~R.~J. 1982b
 The inversion relation method for some two-dimensional 
 exactly solved models in lattice statistics. 
 {\it J. Stat. Phys.} {\bf 28}, 1--41.

 \item[] Baxter, R. J. 1990 Chiral Potts model: 
 eigenvalues of the transfer matrix.
 {\it Phys. Lett. A} {\bf 146}, 110--114.

 \item[] Baxter, R. J. 1991a
 Calculation of the eigenvalues of the transfer matrix of the 
 chiral Potts model. In {\it Proc. Fourth Asia-Pacific Physics 
 Conference}, eds. S.H.~Ahn, Il-T.~Cheon, S.H.~Choh and C.~Lee, 
 Vol. 1, pp. 42--57. Singapore, World-Scientific.


 \item[] Baxter, R. J. 1991b
 Hyperelliptic function parametrization for the 
 chiral Potts model. In {\it Proc. Intnl. Conf. Mathematicians,
 Kyoto 1990}, pp. 1305 --1317. Tokyo, Springer-Verlag.

  \item[] Baxter, R. J. 1993a Elliptic parametrization of the 
 three-state chiral Potts model. In 
 {\it Integrable Quantum Field Theories}, eds. L. Bonora et al, 
 pp. 27--37.  New York, Plenum Press.

  \item[] Baxter, R. J. 1993b Corner transfer matrices of 
 the chiral Potts model. II. The triangular lattice.
  {\it J. Stat. Phys.} {\bf 70}, 535--582.

  \item[] Baxter, R. J. 1998
 Some hyperelliptic function identities that occur in the 
 chiral Potts model. {\it J. Phys. A} {\bf 31}, 6807--6818. 

  \item[] Baxter, R. J. 2003a
 The `inversion relation' method for obtaining the free 
 energy of the chiral Potts model. {\it Physica A} 
 {\bf 322}, 407--431.

 \item[] Baxter, R. J. 2003b
 The Riemann surface of the chiral Potts model free energy 
 function. {\it J. Stat. Phys.} {\bf 112}, 1--26. 

 \item[] Baxter, R.J. 2005a The order parameter of the chiral 
 Potts model. {\it J. Stat. Phys.} {\bf 120}, 1--36.

 \item[] Baxter, R. J. 2005b  Derivation of the order 
 parameter of the chiral Potts model. {\it Phys. Rev. Lett.} 
 {\bf 94}, 130602(3).

 \item[] Baxter, R.J. 2006
 Hyperelliptic parametrization of the generalized order 
 parameter of the chiral Potts model. To appear in {\it The 
 Anziam Journal}.
 
 \item[] Baxter,~R.~J., Perk~J.~H.~H. \& Au-Yang,~H.
 1988 New solutions of the star-triangle relations for the chiral 
 Potts model. {\it Phys. Lett. A} {\bf 128}, 138--142.

 \item[] Howes~S., Kadanoff,~L.~P. \& den Nijs,~M. 1983 Quantum model
 for commensurate-incommensurate transitions.
 {\it Nucl. Phys. B }{\bf 215}[FS7], 169--208.

 \item[] McCoy,~B.~M., Perk,~J.~H.~H. \&  Tang,~S. 1987 Commuting 
 transfer matrices for the four-state self-dual chiral Potts model 
 with a genus-three uniformizing Fermat curve.
 {\it Phys. Lett. A} {\bf 125}, 9--14.

 
  \item[] Onsager,~L. 1944 Crystal statistics. I. A 
 two-dimensional model with an order-disorder transition. 
 {\it  Phys. Rev. } {\bf 65}, 117--49. 

 \item[] von Gehlen,~G. \& Rittenberg,~V. 1985 $Z_n$-symmetric quantum 
 chains with an infinite set of conserved charges and $Z_n$ zero modes.
 {\it Nucl. Phys. B } {\bf 257}[FS14], 351--370.


 \end{description}
 \end{document}